\documentstyle[aps,prl,multicol,epsf]{revtex}
\begin{document}
\draft
\title{\bf Free-carrier screening of polarization fields
in wurtzite GaN/InGaN laser structures}

\author{Fabio Della Sala,$^{1}$ Aldo Di Carlo,$^{1}$ 
 Paolo Lugli,$^{1}$ Fabio Bernardini,$^{2}$ 
Vincenzo Fiorentini,$^{2,3}$
Reinhard Scholz,$^{4}$ and Jean-Marc Jancu$^{5}$}

\address{
\it (1)\, INFM -- Dipartimento di Ingegneria Elettronica, 
Universit\`a di Roma ``Tor Vergata'', Roma, Italy\\
(2)\, INFM -- Dipartimento di Fisica, 
Universit\`a di Cagliari, Italy\\
(3)\, Walter Schottky Institut, TU M\"unchen, Garching, Germany\\
(4)\, Institut f\"ur Physik, TU Chemnitz, Germany\\
(5)\, INFM-Scuola Normale Superiore, Pisa, Italy} 

\date{9 September 1998}

\maketitle

\begin{abstract}
The
free-carrier  screening of macroscopic polarization fields
in wurtzite GaN/InGaN quantum wells lasers
is investigated via a self-consistent tight-binding approach. 
We show that the high carrier 
concentrations found experimentally in nitride 
laser structures effectively screen the built-in
spontaneous and piezoelectric polarization fields, 
thus inducing a ``field-free'' band profile.  
Our results explain some heretofore puzzling experimental data
on nitride lasers, such as the unusually high
lasing excitation thresholds and emission blue-shifts for 
increasing excitation levels.
\end{abstract}


\begin{multicols}{2}
One of the major recent breakthroughs in semiconductor physics
is the realization of blue lasers based on III-V nitride 
heterostructures technology \cite{nakamura:1,nakamura:2}.
In this area, several puzzling issues remain unsettled:
{\it a}) the exact mechanism responsible for laser action is
still a matter for debate \cite{nardelli:1,domen:1}; {\it b})
unusually high threshold densities are required for laser action to
occur  (typically $\sim 10^{13}$ cm$^{-2}$ vs  $\sim 10^{12}$
cm$^{-2}$ for GaAs based devices) \cite{nakamura:2,fang:1,yeo:1};
{\it c}) a blue shift of the transition energy is observed for
increasing excitation powers \cite{chichibu:1,osinski:1}; {\it d}) a
red shift of the transition energy is observed for increasing
well width.

In this Letter, we identify the central reason for these unusual
behaviors, namely the interplay of free-carrier screening and
macroscopic polarization fields. A peculiarity of
 wurtzite nitrides is that of having a non-zero macroscopic polarization,
comprising both a spontaneous and a piezoelectric component 
\cite{bernardini:1}. The  former is a property of low-symmetry
materials  in their ground state,  is independent of strain, and
is absent in zincblende materials (e.g. GaAs). The latter 
appears in the  presence of strain, due to e.g. epitaxy. In III-V
nitrides, both components are very large 
\cite{bernardini:1}; piezoelectric coupling, in particular,
  is orders of magnitude larger that in  e.g. GaAs.   
In nitride  mutilayers,  polarization  manifests itself as  
large ($\sim$ 1 MV/cm) built-in electrostatic fields
\cite{bernardini:2}, produced by  the  charge
 related to polarization changes across heterointerfaces. 
The  field-induced linear bending of the band
edges causes a spatial separation of  confined electrons and holes
within the active layers of the devices \cite{bernardini:2,seo:1}, 
and has therefore important consequences on the optical properties of
nitride-based LED's  or lasers. 

Recent theoretical studied have predicted   a reduced recombination
rate in both GaN/InGaN  \cite{nardelli:1} and 
AlGaN/GaN \cite{honda:1} quantum wells for increasing well width,
due to the field induced separation of electron and holes. Based on
those results, it was argued that  electron-hole
recombination  should be severely hindered, and that
 alternative explanations of lasing, such as e.g. quantum dot
formation, should be sought for. However, the results in question
are not self-consistent, i.e. they neglect the presence of free
electrons and holes in the quantum well. 
Here we show instead that self-consistent free-carrier screening is
the key to understanding the effects of polarization fields on laser
action in nitride devices. 

Specifically, we investigate a   prototypical nitride-based
device, namely a single  In$_{0.2}$Ga$_{0.8}$N quantum well
 pseudomorphically strained to the surrounding GaN barriers. To
account for screening effects, we use the self-consistent
tight-binding (TB) approach \cite{dicarlo} which 
enables us to describe polarization fields, dielectric screening, and
free-carrier screening in a fully self-consistent and
non-perturbative way. TB is used in order to describe  the
electronic structure \cite{dicarlo}  in the whole Brillouin zone up to
several eV's above the fundamental gap,  overcoming all the
natural limitations of simple envelope function  schemes. We use an
$sp^3d^5s^*$ TB   Hamiltonian \cite{scholz:1} representing the
state-of-art of empirical TB; its
 parameters have been determined fitting the DFT-LDA band structure
as outlined in Ref. \onlinecite{scholz:1}. A valence band offset of 0.106 eV
is considered.~\cite{nardelli:1} 
The self-consistent calculation is performed as follows.
The electron and hole quasi-Fermi levels are calculated  for a given
areal charge density ($n_{2D}$) in the quantum well (the {\em sheet density},
 related to  the injection current), and the electron ($n$) and
hole ($p$)  charge distributions in the nanostructure are obtained
\cite{dicarlo}. (The integrated hole and electron charges are equal,
since charge neutrality is assumed). We then solve Poisson equation,
\begin{equation} 
\frac{d}{dz}D =
\frac{d}{dz}\left(-\varepsilon\frac{d}{dz}V+P_T\right)=
e\left(p-n\right).
\label{eq:1}
\end{equation}
The (position-dependent) quantities  $D$,   $\varepsilon$, and $V$,
 are respectively the displacement field, dielectric constant, and
potential. The (position-dependent) transverse
 polarization $P_T$  is the sum of the spontaneous
 component   $P_s$, calculated ab initio \cite{bernardini:1}, and of
 the  piezoelectric component $P_{pz} = 2\, e_{31}\, \epsilon_{xx}+ e_{33}
\,  \epsilon_{zz}$,  involving the calculated \cite{bernardini:1} 
piezoelectric constants  $e_{31}$ and
 $e_{33}$, and the strain tensor components $\epsilon_{xx}$ and
$ \epsilon_{zz}$. These are obtained via  elasticity theory in
 terms of the barrier ($a_b$) and well  ($a_w$) lattice 
parameters~\cite{bernardini:1} and
 elastic constants as $\epsilon_{xx} = (a_b-a_w)/a_w$ and
$\epsilon_{zz} = -2 \epsilon_{xx} C_{13}/C_{33}$.
The constants used in the calculation are listed in Table \ref{tab:1}.
The spontaneous polarization and piezoelectric constants in the alloy
region are obtained by interpolation according to Vegard's rule. The
polarization change across the GaN/InGaN interface is 0.0325 C/m$^2$,
corresponding to a bare interface polarization charge  of
$2\cdot10^{13}$ cm$^{-2}$. The resulting  electrostatic field
in the well, including dielectric screening but no free-carrier screening,
 is 3.25 MV/cm.

Poisson's equation is solved assuming zero field at the boundary. 
The potential thus obtained is plugged into the
TB  Schr\"odinger equation, which is solved to obtain
energies and wavefunctions. The new quasi-Fermi levels are then
calculated, and the procedure is iterated to selfconsistency. 
No analytic approximation is made for the band  dispersions; to
calculate the charge density, a numerical  $k_{\|}$ integration is
performed. Room temperature is assumed throughout all calculations.

Both the sheet carrier density and well width are varied in our
calculations.  This is of interest since it has been shown that the
sheet density  necessary to achieve lasing in InGaN/GaN laser
structures  is unusually large for typical semiconductor lasers
\cite{nakamura:2,fang:1,yeo:1}. The well width is also an   important
parameter in these structures since the total potential drop across
the structure is  directly proportional to the product of polarization
field and well width, if free-carrier  screening is neglected. Indeed,
as mentioned earlier, it has been argued that structures with well
width above than 50 {\AA} should not  be able to produce  laser
action \cite{nardelli:1}.

\begin{figure}
\epsfclipon
\epsfysize=8cm
\centerline{\epsffile{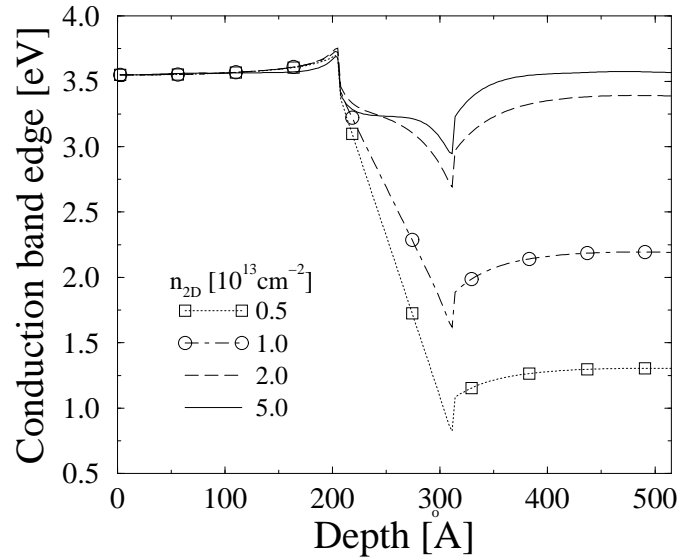}}
\caption{Conduction band edge profile of a 100 {\AA} InGaN quantum
well for several sheet densities}
\label{fig1}
\end{figure}

\begin{figure}
\epsfclipon
\epsfysize=8cm
\centerline{\epsffile{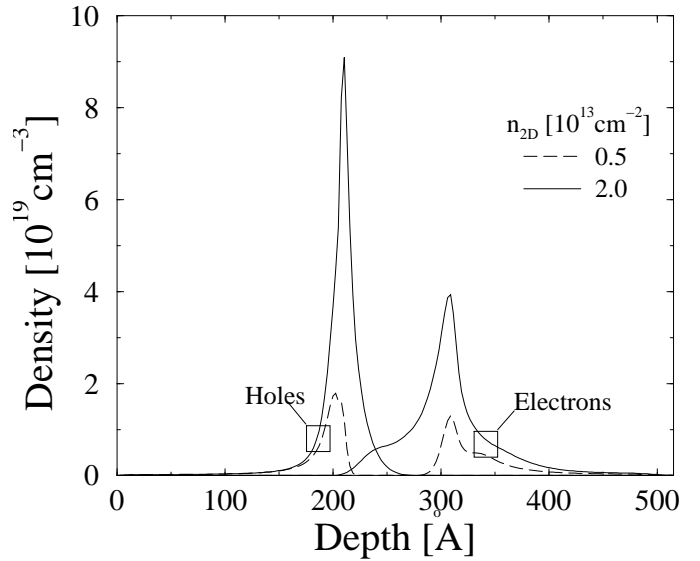}}
\caption{Electron and hole density of the 100  {\AA} well.}
\label{fig2}
\end{figure}

The first result concerns the conduction band profile of a 100 {\AA}
InGaN well, shown in Fig. \ref{fig1} for several sheet densities.
For a relatively substantial sheet density, $n_{2D}$=5 $\cdot$ 10$^{12}$
cm$^{-2}$, a practically uniform electrostatic field as large as 2.5
MV/cm is still present in the well. By increasing the sheet density
(i.e. the  
injection current), free-carrier induced screening becomes more and
more efficient. Upon reaching $n_{2D}$=5 $\cdot$ 10$^{13}$ cm$^{-2}$ a
 quasi--field-free shape of the quantum well is recovered. 
This is easily
understood from the charge densities displayed in Fig.
\ref{fig2} at two typical sheet densities. Electrons and holes are
indeed spatially separated by the polarization field, but
the free-carrier--induced field is opposite to the polarization field.
The two fields tend to cancel each other out for high sheet densities, thus
reestablishing the conditions for electron-hole recombination emission.%

\begin{figure}
\epsfclipon
\epsfysize=10cm
\centerline{\epsffile{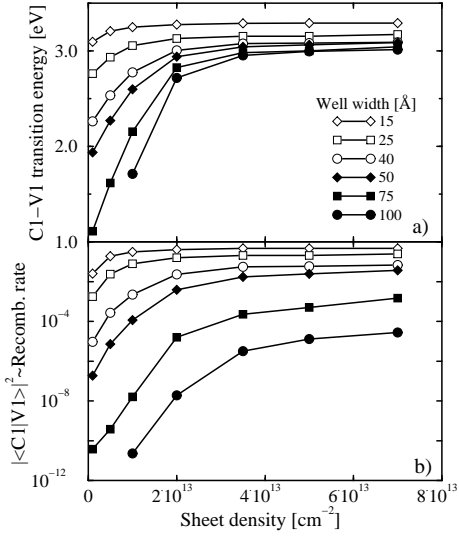}}
\caption{$C1$-$V1$ transition energy and recombination rate vs
 sheet density for several well widths.} 
\label{fig3}
\end{figure}

In Fig. \ref{fig3} we present the calculated
  transition  energy between the highest hole
level ($V1$) and the lowest electron level ($C1$),  and the overlap of
the  $C1$ and $V1$ wavefunctions;   the square of the overlap 
is proportional to the
  recombination rate between the two levels \cite{nota}. 
As seen in Fig. \ref{fig3},
upon increasing the sheet density we observe a blue shift in the 
transition energy and an increased recombination rate.
This is due to the progressive recovery of flat band conditions upon
increasing the sheet density. In addition, the recombination rate
tends to saturate to a finite value at high sheet densities,
and is therefore not vanishingly small for  normal laser operation
conditions (high free-charge density) as previously
claimed \cite{nardelli:1}.
(The difference in recombination rate
at different well widths decreases with increasing sheet density:
 the recombination rate ratio
for widths of 25 {\AA} and 50 {\AA} increases from $3\cdot10^{-4}$ to 0.1 
by increasing the sheet density  from $5 \cdot 10^{12}$
  cm$^{-2}$ to $5 \cdot 10^{13}$ cm$^{-2}$.)

We conclude that laser emission 
can occur via direct electron-hole recombination without having
to invoke composition fluctuations and dot 
formation \cite{nakamura:1,nardelli:1}. However, a rather large sheet
density is necessary to achieve an appreciable recombination rate,
in  agreement with the rather high threshold sheet density observed in
experiments. The  transition blue shift has also been
  observed  experimentally \cite{chichibu:1,osinski:1}. 

\begin{figure}
\epsfclipon
\epsfysize=10cm
\centerline{\epsffile{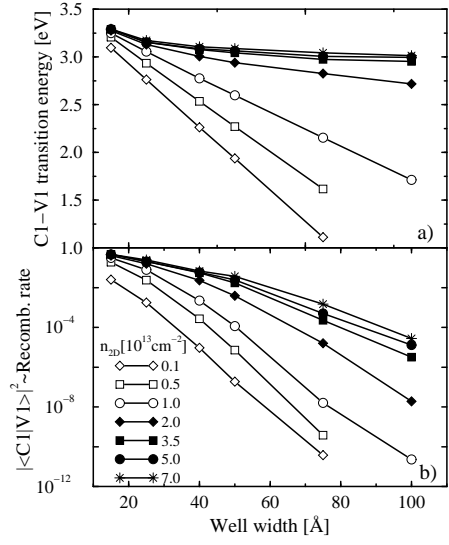}}
\caption{$C1$-$V1$ transition energy and recombination rate vs
 well width for several densities.} 
\label{fig4}
\end{figure}

We notice further that the blue shift just mentioned is relative to the
red-shifted  transition energy in the unscreened polarization field.
This can be read off Fig. \ref{fig4}, displaying transition energies
and recombination rate vs well width.  At low sheet densities,
the  transition  energy suffers a linear red shift for increasing well
width. The transition energy becomes insensitive
to well width (saturating to about 3 eV above 50 {\AA}) only for 
sheet densities  higher than $5 \cdot 10^{13}$  cm$^{-2}$. 
This behavior, also observed experimentally \cite{takeuchi:1}, 
is easily understood by considering again Fig. \ref{fig1}.
At low densities, the polarization field is unscreened and the 
potential drop in the well increases with the well width, thus
reducing the energy separation between $V1$ and $C1$. For high
sheet densities, the field is screened out over most of the well,
with an effective screening length as small as
$\simeq$25 \AA. Since the  V1 and C1
eigenfunctions are localized near the interfaces,  an increase in well
width will not change the transition energy, but will  enhance the
geometrical separation of electrons and holes, eventually  causing a
 reduction in the recombination rate (Fig. \ref{fig4}). However, such
reduction is much smaller than in the unscreened (low sheet density)
case observed  experimentally \cite{seo:1}.

In conclusion, we have shown that free carriers can effectively
screen macroscopic  polarization fields in nitride quantum wells,
resulting in  a non-vanishing recombination rate for large wells
(contrary to previous claims)  in normal  laser operation. A rather
high sheet density is needed to achieve this conditions. We also
explained the red shifts vs well width and blue shifts vs sheet
density as resulting from the interplay of free-carrier screening and
polarization fields.

VF acknowledges support from the Alexander von Humboldt-Stiftung and
from the PAIS program of INFM Section E.

\narrowtext
         
\begin{table}
\begin{center}
\begin{tabular}{ccc}
Parameter             &  GaN      & In$_{0.2}$Ga$_{0.8}$N  \\ \hline
$E_g$ [eV]            &  3.548      & 3.157              \\
$e_{31}$[C/m$^2$]     &  -0.49      &-0.51               \\
$e_{33}$[C/m$^2$]     &  0.73       & 0.77               \\
$C_{13}$[GPa]         &  108        & 105               \\
$C_{33}$[GPa]         &  399        & 359               \\
$P_s$[C/m$^2$]        &  -0.0290    & -0.0296           \\
$\varepsilon_r$       &  10.28      & 11.15             \\
\end{tabular}
\end{center}
\caption{Parameters used in this work.}
\label{tab:1}
\end{table}
     
\end{multicols}

\begin{references}

\bibitem{nakamura:1} S Nakamura {\it et al.},
App. Phys. Lett. {\bf 70}, 2753 (1997);

\bibitem{nakamura:2} 
 S Nakamura {\it et al.},
App. Phys. Lett. {\bf 72}, 2014 (1998); {\it ibid.} {\bf 73}, 832
(1998);  

\bibitem{nardelli:1} M. B. Nardelli, K. Rapcewicz, 
and J. Bernholc, Appl. Phys. Lett. {\bf 71}, 3135 (1997)

\bibitem{domen:1} K. Domen, A. Kuramata, and T. Tanahashi, 
App. Phys. Lett. {\bf 72}, 1359 (1998);  

\bibitem{fang:1} W. Fang, and S. L. Chuang, App. Phys. Lett. 
{\bf 67}, 751 (1995)

\bibitem{yeo:1} Y. C. Yeo, T. C.Chong, M. F. Li, and W.J. Fan, 
J. App. Phys. {\bf 84}, 1813 (1998)

\bibitem{chichibu:1} S. Chichibu, T. Azuhata, T. Sota, and S. Nakamura,
Appl. Phys. Lett. {\bf 69}, 4188 (1996).

\bibitem{osinski:1} M. Osinski {\it et al.},
V. A. Smagley,
 J.  Crystal Growth {\bf 189/190}, 803 (1998)



\bibitem{bernardini:1} F. Bernardini, V. Fiorentini, and D. Vanderbilt, 
Phys. Rev. B {\bf 56}, R10024 (1997).

\bibitem{bernardini:2} F. Bernardini and  V. Fiorentini,
Phys. Rev. B {\bf 57}, R9427 (1997); to be published.

\bibitem{seo:1} Jin Seo Im {\it et al.},
Phys. Rev. B {\bf 57}, R9435 (1998) 

\bibitem{honda:1} T. Honda {\it et al.},
J. Crystal Growth {\bf 189/190}, 644 (1998) 

\bibitem{dicarlo}A. Di Carlo {\it et al.},
Solid State Communications {\bf 98}, 803 (1996); 
A. Di Carlo, MRS Proc. {\bf 491} 389 (1998);

\bibitem{scholz:1} J-M. Jancu, R. Scholz, F. Beltram, and F. Bassani, 
Phys. Rev. B {\bf 57}, 6493 (1998);  R. Scholz, J-M. Jancu, and F. Bassani,
MRS Proc. {\bf 491} 383 (1998).



\bibitem{nota}
We use this envelope-function--like picture (which  can be  relaxed in
the tight-binding scheme, see  M. Graf and P. Vogl, Phys. Rev. B {\bf
51}, 4940 (1995)) in order to make contact with
non-self-consistent data \cite{nardelli:1}.   

\bibitem{takeuchi:1} T. Takeuchi {\it et al.},
Jpn. J. Appl. Phys. {\bf 36},  L382 (1997).


\end{references}
\end{document}